\documentclass[12pt,nohyper]{JHEP3}         
\usepackage{amssymb,amsfonts}
\usepackage{graphicx}

\newcommand{\be}{\begin{equation}}
\newcommand{\ee}{\end{equation}}
\newcommand{\bea}{\begin{eqnarray}}
\newcommand{\eea}{\end{eqnarray}}
\newcommand{\ena}{\end{eqnarray}}
\newcommand{\nn}{\nonumber \\}

\newcommand{\C}{\Gamma}
\newcommand{\e}{\epsilon}
\newcommand{\s}{\sigma}

\def\bfn{\mbox{\boldmath $\nabla$}}

\def\bfn{\mbox{\boldmath $\nabla$}}
\newcommand{\B}{{\cal B}}

\font\mybb=msbm10 at 10pt
\def\bb#1{\hbox{\mybb#1}}

\def\bZ {\bb{Z}}
\def\bR {\bb{R}}

\def\bE {\bb{E}}

\title{The D2 Susy Zoo}

\author{Dongsu Bak\\
        Department of Physics, University of Seoul,\\
Seoul 130-743, Korea\\
        Email: \email{dsbak@mach.uos.ac.kr }}

\author{Nobuyoshi Ohta\\
        Department of Physics, Kinki University,\\
        Higashi-Osaka, Osaka 577-8502, Japan\\
        Email: \email{ohtan@phys.kindai.ac.jp}}

\author{Paul K. Townsend\\
    Department of Applied Mathematics and Theoretical Physics,\\
    Centre for Mathematical Sciences, University of Cambridge,\\
    Wilberforce Road, Cambridge, CB3 0WA, U.K.\\
         Email: \email{p.k.townsend@damtp.cam.ac.uk.}}

\abstract{

We present new supersymmetric solutions of the Dirac-Born-Infeld
equations for time-independent D2-branes, including a 1/2 supersymmetric
`dyonic' D2-brane and various 1/4 supersymmetric configurations that include
`twisted' supertubes, superfunnels with arbitrary planar cross-section,
asymptotically planar D2-branes, and non-singular intersections of
`magnetic' D2-branes. Our analysis is exhaustive for D2-branes in
three space dimensions.}

\preprint{
{\tt hep-th/0612101}$\,,$ \
DAMTP-2006-108, \ KU-TP 010, \ UOSTP-06-104
}

{}

\begin{document}
\section{Introduction}

The low-energy dynamics of D-branes is governed by the Dirac-Born-Infeld (DBI)
action, which generalizes the Dirac brane action to include worldvolume
Born-Infeld (BI) electric and magnetic fields. These fields allow many
more stable configurations than would otherwise be possible since they can
support an otherwise unstable geometry against collapse.
In many cases the stability can be understood as being a consequence of
partial preservation of the supersymmetry of the string theory vacuum.
Although many such `supersymmetric' solutions of the DBI equations have been
found, and their physical implications explored, there has not yet been any
systematic attempt to find {\it all} supersymmetric solutions, in contrast
to the situation for branes without worldvolume fields for which
the mathematics of calibrations allows a complete classification for branes
{\it in vacuo} \cite{Acharya:1998yv,Acharya:1998st}.

Here we initiate a program to classify all time-independent supersymmetric
solutions of the DBI equations for the simplest case of a super D2-brane
in the 10-dimensional Minkowski vacuum of IIA superstring theory (and all
fractions of supersymmetry will refer to fractions of the 32 supersymmetries
preserved by this vacuum). All such solutions have an M-theory interpretation
as supersymmetric (although not necessarily time-independent) M2-branes
\cite{Townsend:1995kk,Townsend:1995af,Schmidhuber:1996fy,Bergshoeff:1996tu}
so it might seem that this case is too simple to yield anything new. However,
the identification of the `extra' space coordinate needed for
$S^1$-compactification to the IIA Minkowski vacuum introduces some subtleties
even for the 1/2 supersymmetric planar D2-branes. Although all such solutions
descend from a planar M2-brane, the 1/2-supersymmetric D2-branes can be
classified according to whether they are `vacuum', `electric', `magnetic'
or `dyonic', and the `dyonic' case is, (as far as we are aware)
a new solution of the DBI equations.

This classification is reminiscent of the classification of intersecting
planar branes in relative motion according to whether the intersection
velocity is subluminal, superluminal or equal to the velocity of
light \cite{Bachas:2002qt}. This similarity is not a coincidence.
The identification of the coordinate of the M-theory circle breaks
the boost invariance in the `extra' direction, leading to a foliation
of the 11-dimensional spacetime by a family of timelike hypersurfaces
that are at rest, in an {\it absolute} sense. One may now consider
the motion of any object in M-theory, such as an M2-brane, with respect
to any one of these rest-frame hypersurfaces, which we call `ether-9-branes',
or `E9-branes', not only because they are `etherial' (in the sense of
having no local physical properties) but also because, collectively, they
play a role analogous to that of the ether in pre-relativistic physics.
Our classification of 1/2 supersymmetric D2-branes corresponds to
the classification of M2-E9 intersections according to the same scheme
as in \cite{Bachas:2002qt}, and the `null' intersection yields the dyonic
1/2 supersymmetric D2-brane.

The possibilities for time-independent supersymmetric D2-branes preserving
less than 1/2 supersymmetry are, of course, much more numerous.
Our analysis is exhaustive only for D2-branes in a 3-dimensional
subspace of the 9-dimensional Euclidean space, for which we find that
all supersymmetric D2-branes are either 1/2 or 1/4 supersymmetric.
Some of the 1/4 supersymmetric solutions are already known. An example
is the original supertube \cite{Mateos:2001qs}, which is
a tubular D2-brane supported by the angular momentum in the BI fields.
Although it is time-independent, it should be considered stationary rather
than static because of the non-zero angular momentum; this feature
is explicit in its IIB superstring-theory dual manifestation as
a ``superhelix'' \cite{Cho:2001ys} and in its M-theory manifestation as
an ``M-ribbon'' \cite{Hyakutake:2002fk}. We find a new tubular solution
that we call a `twisted' supertube because the electric field lines
twist around the tube. A twisted supertube is actually just a supertube
that has been boosted along its axis

The original D2-brane supertube was assumed to have a planar and circular
cross-section, but it was soon realized that other cross-section shapes
are possible \cite{Bak:2001xx,Bak:2001tt,Bak:2002wy}. Indeed, it was shown
in \cite{Mateos:2001pi} that 1/4 supersymmetry allows an arbitrary
cross-sectional curve (which need not even be closed, although the
description as a `tube' becomes inappropriate if the curve is infinite).
This surprising feature is readily understood from the TST-dual manifestation
of the supertube as a ``supercurve'' \cite{Mateos:2002yf}, which is a IIA
string in the T-dual direction carrying a wave with an arbitrary profile.
Given this result, it is natural to wonder whether there exist
supersymmetric `tubular' D2-branes for which the scale of the cross-section
varies along the length of the tube. This possibility was considered
in \cite{Mateos:2001qs}, where it was concluded that any such configuration
would be equivalent to the circularly-symmetric ``dyonic BIon''
of \cite{Gauntlett:2000de}, which was recovered as a `tubular' configuration
with a circular cross-section that varies exponentially along the tube; from
this perspective the term ``superfunnel'' seems more appropriate.
However, the circularly-symmetric superfunnel is found by choosing
a {\it particular} solution of the two-dimensional Laplace equation.
Here we exhibit solutions that yield superfunnels with an {\it arbitrary}
(planar) cross-section. Other solutions of the two-dimensional Laplace
equation yield other 1/4 supersymmetric ``supershapes'' including
asymptotically-planar D2-branes.

We do not attempt a systematic study of D2-brane geometries in Euclidean
spaces of more than three dimensions, but we partially analyze
the conditions for 1/4 supersymmetry. One motivation for this
partial analysis is that the generic cross section for a supertube
in the 9-dimensional space of the 10-dimensional Minkowski IIA vacuum
is known to be an arbitrary curve in the 8-dimensional space transverse
to the supertube `axis' \cite{Mateos:2001pi} and we would like to know
how this result generalizes. Not surprisingly, we find the same result
for the twisted supertubes. In contrast, we find that
the cross-section of a superfunnel is necessarily planar.
Finally, we present a new 1/4 supersymmetric solution for a D2-brane
in a 4-dimensional space that can be interpreted as a (non-singular)
intersection of two, asymptotically planar, `magnetic' D2-branes.

We will begin with a summary of the DBI action for D2-branes,
the supersymmetry preservation condition, and the relation to M2-brane
configurations, thereby collecting together many of the basic formulas
that we will need for the subsequent tour of the ``D2 Susy Zoo''.

\section{Preliminaries}

We choose cartesian coordinates for the 10-dimensional Minkowski metric, such that
\be
ds^2_{10}= -dT^2 + \sum_{I =1}^9(dX^I)^2\, .
\label{cart}
\ee
Let
$\xi^\mu$ ($\mu=0,1,2$) be the D2-brane's worldvolume coordinates.
The induced worldvolume metric is then
\be
ds^2_3 = g_{\mu\nu}\, d\xi^\mu d\xi^\nu \, ,
\ee
where
\be\label{induced}
g_{\mu\nu} = -\partial_\mu T\partial_\nu T
+ \sum_I \partial_\mu X^I\partial_\nu X^I\, .
\ee
The low-energy dynamics for a D2-brane of unit surface tension is
governed by the Dirac-Born-Infeld (DBI) action
\be
I = - \int \!d^3\xi\, \Delta\, ,\qquad \Delta \equiv \sqrt{-\det(g+F)}
\ee
where $F$ is the BI 2-form field strength subject to the Bianchi identity
$dF=0$. Setting $\xi^\mu=(t,\sigma^i)$ ($i=1,2)$, we can write $F$ as
\be\label{fieldstrength}
F=E_i \, dt\wedge d\sigma^i + \B \, d\sigma^1 \wedge d\sigma^2\, ,
\ee
where $E_i$ is the BI electric field, and $\B$ the BI magnetic field density.
The Bianchi identity is
\be
\partial_t \B = \varepsilon^{ij}\partial_iE_j\, .
\ee
Similarly, we can now write the induced metric as
\be
ds^2 = g_{00} \, dt^2 + 2g_{0i}\, dtd\xi^i + h_{ij}\, d\sigma^id\sigma^j
\ee
so that $h_{ij}=g_{ij}$. This allows us to define $h^{ij}$ as the inverse
to $h_{ij}$ (with $g^{ij}$ being the space components of the inverse
to $g_{\mu\nu}$).
After some calculation, one finds that
\be
\Delta^2 = -\det g - \det h\, h^{ij}E_iE_j - g_{00}\, \B^2
- 2\B \, \varepsilon^{ij}E_i\, g_{0j}\, .
\ee

\subsection{Supersymmetry preservation}

Let $(\Gamma_T,\Gamma_I)$ be the (constant) $32\times 32$ spacetime
Dirac matrices, which we may choose to be real. These matrices act on
real $SO(1,9)$ spinors $\epsilon$ that we may decompose as
\be
\epsilon= \epsilon_+ + \epsilon_- \, , \qquad \Gamma_\natural\, \epsilon_\pm
= \pm \epsilon_\pm
\ee
where
\be
\Gamma_\natural \equiv \Gamma_T \Gamma_1 \cdots \Gamma_9\,
\ee
is the 10-dimensional (constant) chirality matrix. The spacetime Dirac
matrices induce reducible ($32\times 32$), and $\xi$-dependent,
worldvolume Dirac matrices $\gamma_\mu$ satisfying
\be
\{\gamma_\mu,\gamma_\nu\} =2g_{\mu\nu}\, .
\ee
We define $\gamma^\mu = g^{\mu\nu}\gamma_\nu$, and similarly for the
antisymmetrized products of worldvolume Dirac matrices; for example
\be
\gamma^{\mu\nu}= g^{\mu\rho}g^{\nu\lambda}\gamma_{\rho\lambda} \, ,\qquad
\left(\gamma_{\mu\nu} \equiv \gamma_{[\mu}\gamma_{\nu]}\right)\, .
\ee
The number of supersymmetries preserved by a given D2-brane configuration
is the dimension of the space of solutions for {\it covariantly-constant}
spinors $\epsilon$ of the equation
\be\label{susypres}
\Gamma\epsilon = \epsilon
\ee
where $\Gamma$ is the `kappa-symmetry' matrix \cite{Bergshoeff:1996tu}
\begin{eqnarray}
\Gamma =
{1\over \Delta}\left[ \gamma_{012} + \varepsilon^{ij}E_i\gamma_j
\Gamma_\natural + \B \gamma_0\Gamma_\natural\right] \, .
\end{eqnarray}
In the cartesian coordinates used here, covariant constancy implies
constancy, so $\epsilon$ must actually be a {\it constant} spinor.
Note that $\Gamma$ is traceless and satisfies the identity
\be
\Gamma^2\equiv 1\, .
\ee
which implies preservation of 1/2 supersymmetry {\it locally}.
However $\Gamma$ is a function of position on the worldvolume, generically,
so the fraction of supersymmetry preserved will generally be less than 1/2.
In fact, generically there will be no non-zero solutions to (\ref{susypres})
so D2-brane configurations preserving any non-zero fraction of supersymmetry
must be special. Finally, note that
\be
\{\Gamma,\Gamma_\natural\} =0\, ,
\ee
which implies that all supersymmetries are broken by a restriction to
chiral 10-dimensional spinors, as expected because there is no supersymmetric
membrane solution of any minimal 10-dimensional supergravity theory.

\subsection{Stationary D2-branes}

Here we are interested in {\it stationary} (time-independent)
supersymmetric D2-brane configurations, so it is convenient to
fix the time-reparametrization invariance by the partial gauge choice
\be
T=t\, .
\ee
We now have a static worldvolume metric with $g_{00}=-1$ and $g_{0i}=0$:
\be
ds^2_3 = - dt^2 + d\sigma^id\sigma^j\, h_{ij}\, ,
\ee
where
\be
h_{ij} = \sum_{I=1}^9 \partial_iX^I\partial_jX^I \, ,
\ee
and hence
\be
\det h = \sum_{I>J} \left({\bfn}X^I \times {\bfn} X^J\right)\left({\bfn}
X^I\times {\bfn}X^J\right)\, ,
\ee
where we use the standard 2D vector calculus notation. We now have
\be\label{deltasquared}
\Delta^2 = (\det h)\left(1-h^{ij}E_iE_j\right) + \B^2\, ,
\ee
where, since $E_i$ and $\B$ are now assumed to be $t$-independent,
the Bianchi identity $dF=0$ reduces, in 2D vector calculus notation, to
\be\label{curlE}
\bfn \times {\bf E} =0\, .
\ee

The induced worldvolume Dirac matrices are
\be
\gamma_0 = \Gamma_T\, ,\qquad \gamma_i = \partial_i X^I\Gamma_I
\ee
so that $\gamma_{012}=\Gamma_T\gamma_{12}$, where
\be
\gamma_{12} = {1\over2}\left({\bfn} X^I\times {\bfn}X^J\right)\Gamma_{IJ}\, .
\ee
The kappa-symmetry matrix $\Gamma$ is
\be\label{redgam}
\Gamma = {1\over\Delta}\left[ {1\over2}\Gamma_T \left({\bfn} X^I \times
{\bfn} X^J\right) \Gamma_{IJ} + \varepsilon^{ij}E_i \partial_j X^I
\Gamma_{I\natural}
+ \B \, \Gamma_{T\natural}\right]\, .
\ee

It is important to appreciate that a stationary supersymmetric D2-brane
will satisfy the DBI equations if and only if the Gauss law constraint,
\be\label{Gauss}
\partial_i {\cal D}^i=0\, ,
\ee
is satisfied, where the electric `displacement' field density is
\be\label{displace}
{\cal D}^i \equiv - {\delta \Delta\over \delta E_i}
= \Delta^{-1}\det h\, h^{ij} E_j\, .
\ee
This follows from consideration of the Hamiltonian formulation;
we will not need this formalism here but we record that the Hamiltonian
density ${\cal H}$ for a stationary D2-brane in the gauge $T=t$ is given by
\be
{\cal H}^2 = \det h + \B^2 + h_{ij}{\cal D}^i {\cal D}^j \left[ 1+
\B^2/\det h \right]\, .
\ee

\subsection{Lift to M-theory}
\label{Mlift}

The D2-brane has an M-theory interpretation as the 11-dimensional supermembrane,
or M2-brane \cite{Townsend:1995kk,Townsend:1995af}. Let $X^\natural$ be the
10th cartesian space coordinate, which becomes the angular coordinate of
the M-theory circle after periodic identification. The unit tension M2-brane
has the action\footnote{We omit the fermions, as they are irrelevant
for the present purposes.}
\be
I_{M2} = -\int d^3\xi \det \left(g^{(M2)}\right)\, .
\ee
The induced worldvolume metric is
\be
g^{(M2)}_{\mu\nu} = g_{\mu\nu} + \partial_\mu X^\natural \partial_\nu X^\natural
\ee
where $g_{\mu\nu}$ is the induced metric of (\ref{induced}). Following the steps
spelled out in detail in \cite{Bergshoeff:1996tu}, one finds that the derivatives
of the M2 worldvolume field $X^\natural(\xi)$ are related to the BI fields of
the D2-brane as follows:
\be
\partial_\mu X^\natural = (\pm)\frac{1}{2\Delta}\,
g_{\mu\nu} \varepsilon^{\nu\lambda\rho} F_{\lambda\rho}\,,
\ee
where $(\pm)$ denotes an appropriate sign corresponding to
the orientation of the D2 embedding.

Let us now specialize to the case of a stationary brane. In the gauge $T=t$, one has
\be
\dot X^\natural = -(\pm)\Delta^{-1}\, \B \, ,\qquad
\partial_i X^\natural = -(\pm)\Delta^{-1}h_{ij}\varepsilon^{jk}E_k\, .
\label{mrelation}
\ee
Equivalently,
\be
\B= -(\pm)\dot X^\natural\, \Delta\, ,\qquad E_i = (\pm)
h_{ij}\, \varepsilon^{jk}\partial_k X^\natural \, (\Delta/\det h)\, .
\ee
Using these expressions in $\Delta$ and solving the resulting equation for
$\Delta$, one finds that
\be
\Delta = \frac{\sqrt{\det h}}{\sqrt{1-\left(\dot X^\natural\right)^2 + h^{ij}
\partial_i X^\natural\partial_j X^\natural}}\, .
\ee
Note that the quantity
\be
-\frac{1}{2}F^{\mu\nu}F_{\mu\nu} = 
h^{ij}E_iE_j -\left(\B^2/\det h\right) = \left[ h^{ij} \partial_i X^\natural
\partial_j X^\natural -\left(\dot X^\natural\right)^2\right]
\left(\Delta^2/\det h\right)\, ,
\ee
is invariant with respect to the $Sl(2;\bR)$ worldvolume Lorentz group.

\section{One-half supersymmetry}
\label{sec:half}

Before considering configurations preserving less than 1/2 supersymmetry,
we consider the condition for 1/2 supersymmetry, which is
\be\label{onehalfsusy}
\left[{1\over2}\Gamma_T \left({\bfn} X^I \times {\bfn} X^J\right) \Gamma_{IJ}
+ \varepsilon^{ij}E_i \partial_j X^I \Gamma_{I\natural}
+ \B \, \Gamma_{T\natural} - \Delta\right]\epsilon=0
\ee
As $\epsilon$ is constant, this condition will imply preservation of
1/2 supersymmetry only if all terms are proportional to $\Delta$.
In particular we require
\be
{\bfn} X^I \times {\bfn} X^J = \Omega^{IJ}\, \Delta
\ee
for some constant antisymmetric $9\times 9$ matrix $\Omega^{IJ}$. Fixing
the worldspace diffeomorphisms by choosing $X^1=\sigma^1,X^2=\sigma^2$
we learn that $\Delta$ is constant in this gauge, in which case all space
components $X^I$ must be linear in the worldspace coordinates $\sigma^i$
in order that $ {\bfn} X^I \times {\bfn} X^J$ be constant for all $(I,J)$.
This implies that the brane geometry is a plane, with the constants
${\bfn} X^I \times {\bfn} X^J$ being proportional to the projections of
an area element of the brane onto the $I-J$ plane. Clearly, we may orient
the planar brane such that the only non-zero entries of $\Omega$ are
$\Omega^{12}=-\Omega^{21} = 1/\Delta$. We then have $h_{ij}=\delta_{ij}$,
and hence
\be
\Delta^2 = 1+ \B^2 - |{\bf E}|^2\, .
\ee

\subsection{Classification}

Now that we know that $\Delta$ is constant, the requirement that all
coefficients in (\ref{onehalfsusy}) be proportional to $\Delta$ implies that
\be
{\bf E}={\bf n}\, E\, ,\qquad \B=B\, ,
\ee
for fixed unit 2-vector ${\bf n}$ and {\it constants} $E$ and $B$.
The supersymmetry preservation condition now reduces to
\be
\left[\Gamma_T\Gamma_{12} +\varepsilon^{ij}E\, n_i \Gamma_{j\natural}
+B \Gamma_{T\natural} - \Delta\right] \epsilon=0\, ,
\ee
where
\be
\Delta = \sqrt{1-E^2+B^2}\, .
\ee
This condition can be satisfied for any constants $(E,B)$ provided that
$E<\sqrt{1+B^2}$, and so all such planar D2-brane configurations
preserve 1/2 supersymmetry.

The energy density is
\be
{\cal H} = \frac{1+B^2}{\sqrt{1+B^2-E^2}} \ge1\, ,
\ee
with equality for the `vacuum' D2-brane with $E=B=0$, which is the only
1/2 supersymmetric configuration that is invariant under the $SO(1,2)$
worldvolume Lorenz group.
All other 1/2 supersymmetric static D2-branes break the worldvolume
Lorentz invariance, and as $E^2-B^2$ is a Lorentz invariant there are
three cases to consider:

\begin{itemize}

\item $E^2-B^2<0$. In this case we may boost to a frame in which $E=0$.
In this frame the non-zero magnetic density breaks $SO(1,2)$ to $SO(2)$,
the worldspace rotation group. Because the boost invariance is broken,
a boost generates a new 1/2 supersymmetric D2-brane with electric
as well as magnetic BI field.

\item $E^2-B^2>0$. In this case we may boost to a frame in which $B=0$.
The non-zero electric field in this frame, which is constrained by
$|{\bf E}|<1$, breaks $SO(1,2)$ to $SO(1,1)$, which is the group of
boosts in the direction of the electric field. A boost in the orthogonal
direction generates a new 1/2 supersymmetric D2-brane with a magnetic field.

\item $E=B \ne0$. This case is intrinsically `dyonic' in the sense that
there is no frame for which either the electric or the magnetic field
is zero. We now have $\Delta=1$ and
\be
\Gamma = \Gamma_T\Gamma_{12} + B\left(\Gamma_T+\Gamma_2\right)\Gamma_\natural\, .
\ee

\end{itemize}

We may summarize this state of affairs by saying that a non-vacuum 1/2
supersymmetric D2-brane is `magnetic', `electric', or `dyonic' according to
whether the 3-vector $(B,E_1,E_2)$ is, respectively, timelike, spacelike or null.

\subsection{M-theory and E-branes}

Each of the possible 1/2 supersymmetric D2-branes must lift to a planar
M2-brane in the $\bE^{(1,9)}\times S^1$ vacuum of M-theory, and it is
instructive to see how the distinction between magnetic, electric and dyonic
D2-branes arises in this context. As in subsection \ref{Mlift}, we let
$X^\natural$ be the coordinate of the M-theory circle. We may assume that
the plane of M2-brane is spanned by a vector along the 1-axis, say, and
another vector in the $2-\natural$ plane. Taking into account that
$(X^1,X^2)=(\sigma^1,\sigma^2)$, this implies that $X^\natural$ is linear
in $\sigma^2$. Allowing, too, for a linear time-dependence of $X^\natural$,
we have
\be
\dot X^\natural =u \, ,\qquad \partial_2X^\natural = \tan\theta\, ,
\ee
for constant $u$ and constant angle $\theta$, which is the angle that
the M2-brane makes to the $X^2$-axis. From subsection~\ref{Mlift},
we read off the corresponding BI fields:
\be
E= E_1 = \frac{\sin\theta}{\sqrt{1-v^2}}\, ,\qquad B= -\frac{v}{\sqrt{1-v^2}}\, ,
\ee
where
\be
v= u\cos\theta
\ee
is the physical transverse velocity of the M2-brane. It follows that
\be
E^2-B^2 = \frac{\sin^2\theta \left(1-v_{int}^2\right)}{1-v^2}\, ,
\ee
where we have defined
\be
v_{int} \equiv u/\tan\theta = v/\sin\theta\, .
\ee
We see from this result that the D2-brane will be `electric' or `magnetic'
according to whether the velocity $v_{int}$ is subluminal or superluminal,
and the dyonic D2-brane corresponds to $v_{int}=1$.
But what is the intrinsic significance of $v_{int}$?

To answer this question, we begin by recalling a discussion of Bachas and Hull
on intersecting branes \cite{Bachas:2002qt}. If the velocity of the intersection
is $v_{int}$ then one must distinguish between the three cases $v_{int}<1$,
$v_{int}=1$ and $v_{int}>1$.
For $v_{int}<1$ one can boost to a frame in which the intersection is at rest,
and in this frame one has static intersecting branes. For $v_{int}>1$ one can boost
to a frame in which the branes are parallel but in relative motion.
It seems as though a similar analysis should be applicable here but, if so,
where is the `other' brane with respect to which the M2-brane is in motion?

To answer {\it this} question, we recall that the identification of the
$X^\natural$ coordinate breaks the 11-dimensional Lorentz invariance;
in particular it breaks the invariance under boosts in the $X^\natural$
direction, thereby introducing globally-defined rest-frames for motion
in this direction. Specifically, any hypersurface of constant $X^\natural$
is at rest, and can be viewed as a kind of `etherial' 9-brane. Recalling that
hypothetical material defining the absolute rest frame in pre-relativistic
physics was called the `ether', we propose to call this
an `ether' 9-brane, or `E9-brane'. An E9-brane has no local physical
properties, but is nevertheless a convenient way of thinking about
the globally-defined rest frames implied by the existence of the M-theory
circle. It is convenient because we can now apply a Bachas-Hull-type argument
to the M2-E9 intersection. The velocity of this intersection is precisely
$v_{int}$, so we get a classification of D2-branes according to whether
$v_{int}$ is less than, greater than or equal to the velocity of light.
As we have just shown, this classification coincides with the
`intrinsic' classification into electric, magnetic and dyonic D2-branes.

\section{D2 in 3D}
\label{sec:original}

In the following section, we will present a systematic analysis that uncovers
all possible time-independent supersymmetric D2-branes for which the D2-brane
geometry is a surface in Euclidean 3-space with coordinates $(X^1,X^2,X^3)$.
Many special cases can be found by non-systematic methods of course,
and we are first going to present such an analysis, based on the original
approach in \cite{Mateos:2001qs}. We do this partly because we wish to correct
an error in \cite{Mateos:2001qs} which led to a puzzle that we resolve here,
and also because it turns out that we do actually find all the 1/4
supersymmetric solutions this way.

Following \cite{Mateos:2001qs}, we introduce the cylindrical polar
coordinates $(R,\Phi,X)$:
\be\label{cylindrical}
X^1= R\cos \Phi\, , \qquad X^2 = R\sin \Phi\, , \qquad X^3=Z\, .
\ee
For convenience of comparison with \cite{Mateos:2001qs}, we also
relabel the worldspace coordinates
\be\label{tubular}
\sigma^1 = z\, , \qquad \sigma^2=\varphi \, ,
\ee
with $\varphi$ an angular coordinate (such that $\varphi\sim \varphi + 2\pi$).
This allows us to fix the worldspace parametrization invariance by the gauge choice
\be
Z=z\, ,\qquad \Phi =\varphi\, .
\ee
The induced worldspace line-element is
\be
h_{ij}d\sigma^i d\sigma^j = \left(1+ R_z^2\right) dz^2
+ 2R_z R_\varphi dz d\varphi + \left(R^2+ R_\varphi^2\right) d\varphi^2\, ,
\ee
where $R_z=\partial_z R$ and $R_\varphi= \partial_\varphi R$. In this approach,
the cross-section of the tube is assumed to be planar from the outset, but
since $R$ is, a priori, a function of both $z$ and $\varphi$, allowance is
made for a possible change of scale along the axis of the tube (which leads
to what we will here call `superfunnels') and an arbitrary planar shape
(although only the circular supertube was actually found in \cite{Mateos:2001qs}).
We write the BI 2-form as
\be
F= E_z\, dt\wedge dz + E_\varphi \, dt\wedge d\varphi + \B \, dz\wedge d\varphi\, .
\ee
Note the allowance of an electric field component $E_\varphi$ around
the tube as well as the component $E_z$ along it. In this respect we have
a generalization of the analysis of \cite{Mateos:2001qs}.

A calculation now yields
\be
\Delta^2 = \left(R^2+ R_\varphi^2\right) \left(1-E_z^2\right) + \B^2 + R^2R_z^2
-E_\varphi^2 \left(1+ R_z^2\right) + 2R_zR_\varphi E_z E_\varphi\, ,
\ee
from which we compute that
\bea
{\cal D}_z &=& \Delta^{-1}\left[\left(R^2+ R_\varphi^2\right)E_z
-R_zR_\varphi E_\varphi\right]\, ,\nonumber \\
{\cal D}_\varphi &=& \Delta^{-1}\left[\left(1+ R_z^2\right)E_\varphi
-R_zR_\varphi E_x\right] \, .
\eea
The Hamiltonian density is
\be
{\cal H}= \Xi^{-1}\sqrt{\left[\Xi^2 + |{\cal D}|^2\right] \left[\Xi^2 + \B^2\right]}
\ee
where
\bea
\Xi^2 &\equiv& R^2\left(1+R_z^2\right) + R_\varphi^2\, ,\nonumber\\
|{\cal D}|^2 &=& \left(1+R_z^2\right){\cal D}_z^2 + 2R_zR_\varphi
{\cal D}_z{\cal D}_\varphi + \left(R^2+R_\varphi^2\right){\cal D}_\varphi^2\, .
\eea

The induced worldvolume Dirac matrices are
\be
\gamma_0= \Gamma_T\, ,\qquad \gamma_z = \Gamma_Z + R_z \Gamma_R\, ,\qquad
\gamma_\varphi = R\Gamma_{\underline \Phi} + R_\varphi \Gamma_R
\ee
where $\Gamma_{\underline \Phi}$ is the Dirac matrix in the obvious
orthonormal frame; it is constant and squares to the identity
matrix. Following \cite{Mateos:2001qs}, we note that a covariantly
constant spinor in the new cylindrical polar coordinates is not constant
but instead takes the form
\be
\epsilon =\exp\left( {1\over2}\Phi \Gamma_{R{\underline\Phi}}\right)\,
\epsilon_0 \, .
\ee
Following the steps sketched in \cite{Mateos:2001qs}, and taking into
account that $\Phi=\varphi$, we now find that the supersymmetry
preservation condition can be put into the form
\bea\label{MTcondition}
0&=&\left[RR_z \Gamma_{TR{\underline\Phi}}
+ \B \Gamma_{T\natural} - E_\varphi\Gamma_{Z\natural} - \Delta\right]
\epsilon_0\nonumber\\
&& + \ \exp\left(-\varphi \Gamma_{R{\underline\Phi}}\right)
\left[\gamma_\varphi \Gamma_\natural\left(\Gamma_{TZ\natural} + E_z\right)
- E_\varphi R_z\Gamma_{R\natural}\right] \epsilon_0\, .
\eea
For $E_\varphi=0$, this should reduce to the condition found
in \cite{Mateos:2001qs} but it does not quite do so. The first set of
bracketed terms in eq. (20) of \cite{Mateos:2001qs} erroneously includes
an additional $R_z R_\varphi\Gamma_T$ term that led the authors to conclude
that 1/4 supersymmetry requires either $R_z=0$ or $R_\varphi=0$. Since
only such configurations were then considered, this error had no further
consequences in \cite{Mateos:2001qs} but we must now re-analyse
the possibilities, which we do taking into account the additional
possibility of non-zero $E_\varphi$.

If one assumes that the worldvolume fields are $\varphi$-independent, then,
as argued in \cite{Mateos:2001qs}, the two square-bracketed terms on
the right hand side of (\ref{MTcondition}) must vanish independently.
This requirement is not obviously necessary when the worldvolume fields are
allowed to depend on $\varphi$ but, as will be seen in the following section,
relaxation of it does not yield any new solutions. We therefore proceed
by assuming that each of the two bracketed terms is zero. Observing that
the gamma-matrices in the first two terms of the second bracket square to
the identity whereas the gamma-matrix in the last term squares to minus
the identity, we deduce that the terms of the second bracket vanish if and only if
\be
\label{dichotomy}
E_z = \pm 1\, , \qquad E_\varphi R_z=0\, ,
\ee
and
\be\label{TZnat}
\Gamma_{TZ\natural}\, \epsilon= \mp \epsilon\, .
\ee
The supersymmetry preservation condition then becomes
\be
\label{MTcondition2}
\left[RR_z\, \Gamma_{TR{\underline\Phi}} + \B\, \Gamma_{T\natural}
- E_\varphi\, \Gamma_{Z\natural} -\Delta \right] \epsilon_0 = 0
\ee
where, now,
\be
\Delta^2 = \B^2 + \left(RR_z\right)^2 -E_\varphi^2\, .
\label{deleq}
\ee

We see from (\ref{dichotomy}) that there are two alternatives:
either (i) $R_z=0$ or (ii) $E_\varphi=0$.
We will consider them in turn.

\subsection{Supertubes}
\label{sec:supert}

For the choice
\be
R_z=0\, ,
\ee
the supersymmetry preservation condition becomes
\be
\left(\B\Gamma_{T\natural} + E_\varphi \Gamma_{Z\natural}\right)\epsilon =
\sqrt{\B^2 -E_\varphi^2}\ \epsilon\, .
\ee
Since both $\Gamma_{T\natural}$ and $\Gamma_{Z\natural}$ commute with
$\Gamma_{TZ\natural}$, this condition is compatible with (\ref{TZnat}),
but the two together imply preservation of 1/4 supersymmetry only if
\be\label{EBvarphi}
E_\varphi = \beta\B\, ,\qquad \beta^2<1
\ee
for some constant $\beta$. The Gauss law reduces to $\partial_z\B=0$
in this case, consistent with invariance under translations along the $Z$-axis.
The standard supertube, with an arbitrary planar cross-section,
is found by setting $E_\varphi=0$. Details of the $E_\varphi\neq0$ case
will be left until our later discussion allowing for a non-planar cross-section.

We now have a 1/4 supersymmetric tubular configuration determined by
the arbitrary functions $R(\varphi)$ and $\B(\varphi)$. To facilitate comparison
with the more general solutions that we present in section~\ref{sec:general},
we recall here that we fixed the worldspace diffeomorphism by setting $Z=z$ and
$\Phi=\varphi$, so that the functions $X^1(\varphi)$ and $X^2(\varphi)$
are not independent because they are both determined by the function
$R(\varphi)$. However, we could now suppose that $\varphi$ is a function of
some new angular variable $\psi$, in which case
\be
\B\, dz\wedge d\varphi = B\, dz \wedge d\psi
\ee
where $B= \B(\partial \varphi/\partial\psi)$ is a {\it constant}.
In this reparametrization of the solution, it is now the two functions
$X^1(\psi)$ and $X^2(\psi)$ that are independent, and they determine
an arbitrary curve in the $X^1$-$X^2$ plane.

\subsection{Superfunnels}
\label{subsec:sf}

We now suppose that $R_z\neq0$, in which case we must set
\be
E_\varphi=0\, .
\ee
The supersymmetry preservation condition is now
\be
\left(RR_z \Gamma_{TR{\underline\Phi}} + \B\Gamma_{T\natural}\right)\epsilon =
\sqrt{\B^2 + R^2R_z^2}\, \epsilon\, .
\ee
Since both $\Gamma_{TR{\underline\Phi}}$ and $\Gamma_{T\natural}$ commute
with $\Gamma_{TZ\natural}$, this condition is compatible with (\ref{TZnat}),
but the two together imply preservation of 1/4 supersymmetry only if
\be
\B = B_0 RR_z
\ee
for some constant $B_0$. The Gauss law constraint in this case is
\be\label{GaussR}
\partial_z \left[ \frac{R}{R_z} + \frac{R_\varphi^2}{RR_z}\right]
= \partial_\varphi^2 \log R\, .
\ee
To see the significance of this constraint, we switch independent
variables from $(z,\varphi)$ to $(\rho\equiv R, \varphi)$, taking
$z$, which equals $Z$, as the dependent variable, and use the relations
\be
\left(\partial_z\right)_\varphi = {1\over \left(Z_\rho\right)_\varphi}\,
\left(\partial_\rho\right)_\varphi \, , \qquad
\left(\partial_\varphi\right)_z = \left(\partial_\varphi\right)_\rho
- {\left(Z_\varphi\right)_\rho \over
\left(Z_\rho\right)_\varphi } \, \left(\partial_\rho\right)_\varphi\, .
\ee
We find that
\be
R_z = 1/Z_\rho\, ,\qquad R_\varphi = - Z_\varphi/Z_\rho\, ,
\ee
where it should be clear which variables are being kept fixed from
the convention that lower/upper case variables are independent/dependent.
One then finds that (\ref{GaussR}) is equivalent to
\be\label{laplace}
\nabla^2 Z =0\, .
\ee
In other words $Z$ is a solution of the 2D Laplace equation. Note too that
\be
F= \pm\, dt\wedge dz + \B dz\wedge d\varphi
= \pm (\partial_{X^i} Z )\, dt\wedge dX^i + B_0 dX^1\wedge X^2\, .
\ee

The general periodic solution of the Laplace equation is
\be\label{general}
Z = Z_0 + Q \log \rho + {\sum_{k\in \bZ}}' \left[\left(c_k \cos (k\varphi) +
\tilde c_k \sin (k\varphi)\right)\right]\rho^k
\ee
for constants $Z_0$, $Q$ and $(c_k, \tilde c_k)$, where the prime on the
sum indicates that the $k=0$ term is omitted. It was noted
in \cite{Mateos:2001qs} that the particular solution $Z=Z_0 + Q\log \rho$
is equivalent to the dyonic BIon of \cite{Gauntlett:2000de}.
More generally,
\be
\oint d\varphi\, Z \propto \log \rho\, .
\ee
for any solution of the Laplace equation with non-zero $Q$.
This is the expected logarithmic bending of a D2-brane due to a charge $Q$,
which can be interpreted as the charged endpoint of an infinite IIA string.
Equivalently, we have a IIA string that has been `blown-up' to a funnel-shaped
D2-brane, with a planar cross-section of arbitrary shape that grows
exponentially along the axis of the funnel, at least `on average'. Since they
preserve 1/4 supersymmetry we call them `superfunnels'.

\subsection{Other Supershapes}

When $Q=0$, we have a new type of 1/4 supersymmetric D2-brane that is neither
a supertube nor a superfunnel, arising from the multipole expansion of $Z$.
A simple example, which might be called a ``super-dipole'' is
\be
Z= \cos\varphi /\rho\, .
\label{exa}
\ee
This has the feature that $Z\to 0$ as $\rho\to\infty$, so the brane geometry is
asymptotically planar. We depict the shape of the surface in Figure~1.
One finds that $\B=B_0 R R_z = -B_0\rho^3/\cos\varphi$ for this configuration,
which appears to blow up for $\varphi=\pi/2$ and as $\rho\to\infty$, but this
is an artifact of the choice of coordinates. As we showed above, $\B$ is
proportional to the area element and hence constant in local cartesian coordinates.

The projections on the spinor $\epsilon$ required by the 1/4 supersymmetry
of all these $Q=0$ solutions are exactly the same as for the superfunnel
solutions, which suggests an interpretation as a IIA string that passes
through a planar D2-brane. If the point of intersection of such a string
is split, one has two string ends of opposite charge. Such a zero charge
distribution would have multipole moments of all orders, so the pure dipole
solution of (\ref{exa}) must correspond to a much smoother charge distribution
of zero net charge, which is possibly why
it does not have the appearance of a string intersecting a D2-brane.

\begin{figure}[ht!]
\begin{center}
\includegraphics[totalheight=7cm]{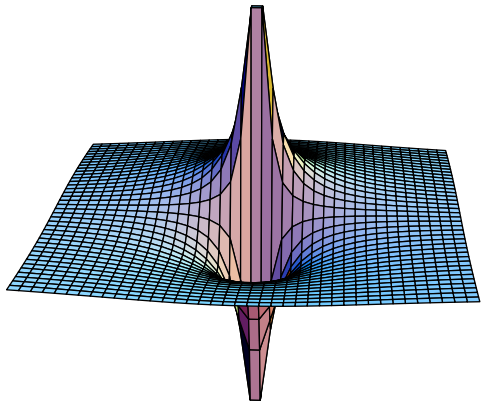}
\end{center}
\vspace{-20mm}
\caption{\small
The simple example~(\ref{exa}) that is asymptotically planar.}
\end{figure}

Let us consider the M-theory lift of the ``super-multipole'' solutions.
Taking into account a sign associated to the orientation of D2 embedding, one has
\be
(\pm)\Delta = \sqrt{1+ B_0^2} \, R R_z=
\sqrt{1+B_0^2} \frac{\cos^2 \varphi}{z^3}\,.
\ee
{}From (\ref{mrelation}) we see that the velocity in the $\natural$ direction is
\be
\dot{X}^\natural = \frac{B_0}{\sqrt{1+B_0^2}}\,.
\ee
In fact, this result also holds for any of the superfunnel
configurations, and it implies that both superfunnels and super-multipoles
are moving uniformly with constant velocity along the M-theory circle.

Since $E_\varphi=0$, the second equation of (\ref{mrelation}) becomes,
for the configuration defined by (\ref{exa}),
\be
\partial_z {X}^\natural =\frac{\tan \varphi}{\sqrt{1+B_0^2}}\,,
\qquad
\partial_\varphi {X}^\natural =\frac{z}{\sqrt{1+B_0^2}\cos^2 \varphi}\, .
\ee
These equations can be integrated straightforwardly:
\be
X^\natural =\frac{1}{\sqrt{1+B_0^2}} \Bigl(
B_0 t + z \tan \varphi\Bigr)\,,
\ee
from which one may notice that $X^\natural$ is also a harmonic function
satisfying $\nabla^2 X^\natural=0$ in the $\rho$ and $\varphi$ space.
This is actually a general result that is true of all superfunnels and
super-multipole configurations.

\section{Systematics}
\label{Systematic}

We now present a systematic analysis of 1/4 supersymmetric D2-branes in
three Euclidean space dimensions. For this purpose it is convenient to revert
to the cartesian space coordinates $X^a; a=1,2,3$, and worldvolume coordinates
$(t,\sigma^1,\sigma^2)$ and fix only the time reparametrizations by the
usual gauge choice $T=t$. We assume that the worldvolume fields $(X^a,{\bf E},\B)$
are $t$-independent.

Because of the restriction to a 3-dimensional space, we have
\bea
\Delta^2 = \sum_a \left(G_a G_a -H^a H^a\right) +\B^2.
\label{delta2}
\ena
where
\be
G_a ={1\over 2} \epsilon_{abc}\bfn X^b \times \bfn X^c\, ,\qquad
H^a ={\bf E} \times\bfn X^a\, .
\label{gh}
\ee
In this notation, the supersymmetry preservation condition becomes
\be
\left[H^a \Gamma_{a\natural} +{1\over 2}\epsilon^{abc} G_a \Gamma_T\Gamma_{bc}
+\B\, \Gamma_{T\natural} -\Delta \right] \epsilon =0\, .
\label{ks6}
\ee

Because the spacetime is effectively assumed to be 4-dimensional, the results
we obtain must be the same as if we had started from the super-D2-brane
in this spacetime dimension. Consequently, we may interpret $\epsilon$ as
a complex 4-component Dirac spinor\footnote{Supersymmetry requires the spinor
parameter $\epsilon$ to be a Dirac spinor.One way to see this is to observe
that the four-dimensional super-D2-brane
action is equivalent to the action of the 5-dimensional supermembrane
in exactly the same way as the string theory super-D2-brane action is
equivalent to the action of the 11-dimensional supermembrane.} and
the matrices $(\Gamma_T,\Gamma_a; a=1,2,3)$
as standard four-dimensional $4\times 4$ Dirac matrices and $-i\Gamma_\natural$
as their product. A convenient representation is
\bea
\C_T = \left(\begin{array}{cc}
0 & {\bf 1} \\
-{\bf 1}\  & 0
\end{array}\right)
\, , \qquad
\C_a = \left(\begin{array}{cc}
0 & \s_a \\
\s_a & 0
\end{array}\right),
\ena
in which case
\be
\Gamma_\natural = i \Gamma_T\Gamma_{123} = \left(\begin{array}{cc}
-{\bf 1}\ & 0 \\
0 & {\bf 1}
\end{array}\right),
\ee
where {\bf 1} is a $2 \times 2$ unit matrix.
The supersymmetry preservation condition is now
\small{
\bea\label{matrixeq}
\left(\begin{array}{cc}
-\Delta& a_0 + \sum_a a_a\s_a \\
a_0 -\sum_a a_a\s_a & -\Delta
\end{array}
\right) \left( \begin{array}{c}
\e_+ \\
\e_-
\end{array} \right)
=0\; ,
\ena
}
where $\e_\pm$ are two-component $SU(2)$-spinors spanning the $\mp$ eigenspaces of
$\Gamma_\natural$, and
\be
a_0=\B\, ,\qquad a_a = H^a + iG_a\, .
\ee
Note the manifest $SU(2)$ invariance of the equation (\ref{matrixeq}),
and its invariance under a common $U(1)$ phase rotation of $\e_\pm$.
This equation is equivalent to
\bea\label{matrix2}
\left( \begin{array}{c}
\e_3 \\
\e_4
\end{array} \right) = \frac{1}{\Delta}
\left(\begin{array}{cc}
a_0-a_3 & -a_1 +ia_2 \\
-a_1-ia_2 & a_0+a_3 \\
\end{array}
\right)
\left( \begin{array}{c}
\e_1 \\
\e_2
\end{array} \right)\, .
\ena
where $(\e_1,\e_2)$ are the components of $\e_+$ and $(\e_3,\e_4)$ are
the components of $\e_-$. Observe that $\sum_a H^aG_a =0$, and hence that
\be\label{delHG}
\Delta^2 = a_0^2 - \sum_a a_a a_a\, .
\ee

Given that $(\e_1,\e_2)$ are arbitrary complex constants, we can show that
$(\e_3,\e_4)$ are also constants if and only if $(a_0,a_1,a_2,a_3)$,
and hence $\Delta$, are constants. This is the case of 1/2 supersymmetry.
Our results of section \ref{sec:half} are recovered by fixing the
worldspace diffeomorphism invariance appropriately.
Preservation of any fraction of supersymmetry less than 1/2 is possible only
if $\e_1$,$\e_2$ and their complex conjugates are subject to linear relations.
We may use symmetries of (\ref{matrix2}) to bring any such relations into
a `standard' form using the $SU(2)$ rotational invariance and $U(1)$ phase
invariance. Given a single real linear relation between $(\e_1,\e_2)$
and their complex conjugates, we may use the phase invariance to arrange for
$\e_2$ to be real. This leads to configurations that preserve at least
3/8 supersymmetry. However, the conditions for $\e_3$ and $\e_4$ to be
constant remain the same as they were before so all configurations
preserving at least 3/8 supersymmetry actually preserve 1/2 supersymmetry.

We learn from this that to find configurations preserving less than
1/2 supersymmetry we need to impose at least two real linear relations
on $\e_1$, $\e_2$ and their complex conjugates. Given two real relations
we will find configurations preserving at least 1/4 supersymmetry (we say
``at least'' because the 1/2 supersymmetric configurations will be
included). Given three real relations we will find configurations preserving
at least 1/8 supersymmetry. Any two real linear relations are equivalent
to one complex linear relation on $\e_1,\e_2$. We may now use the $SU(2)$
symmetry to arrange for this relation to be $\e_2=0$. In this case,
(\ref{matrix2}) implies
\be
\label{3412}
\epsilon_3 = \Delta^{-1}\left(a_0-a_3\right)\epsilon_1\, ,\qquad
\epsilon_4 = - \Delta^{-1}\left(a_1+ia_2\right)\epsilon_1\, ,
\ee
In the case that there is one more linear relation, which will now be
a relation between the real and imaginary parts of $\e_1$ we can use
the $U(1)$ invariance to arrange for this relation to be ${\cal I}m \e_1=0$,
so that $\e_1$ is real. In this case (\ref{matrix2}) again implies
(\ref{3412}) but with $\e_1$ now restricted to be real.
However, irrespective of whether $\e_1$ is real or complex, constancy
of $(\e_3,\e_4)$ requires both $(a_1+ia_2)/\Delta$ and $(a_0-a_3)/\Delta$ to
be constant. It follows that all configurations preserving at least
1/8 supersymmetry actually preserve at least 1/4 supersymmetry, and
that the conditions for this are
\bea
\label{onetwo}
{\cal B} - {\bf E}\times \bfn X^3 + i \bfn X^1 \times \bfn X^2
&\propto& \Delta\, ,\nonumber \\
\left({\bf E} -\bfn X^3\right) \times \left(\bfn X^1 +i \bfn X^2 \right)
&\propto& \Delta \, .
\eea
{}From the imaginary part of the first of these equations we see that
\be
\bfn X^1 \times \bfn X^2 = {\cal A}\, \Delta\, ,
\ee
for some {\it constant} ${\cal A}$, which determines the projection of
an area element of the D2-brane onto the $X^1$-$X^2$ plane. We must now
consider separately the ${\cal A}=0$ and ${\cal A}\ne0$ cases.

\subsection{${\cal A}=0$}

When ${\cal A}=0$, it is convenient to revert to the cylindrical 3-space
coordinates of (\ref{cylindrical}) and the worldspace coordinate notation
of (\ref{tubular}). We may then fix the worldspace diffeomorphism
invariance by the gauge choice $Z= z$ and $\Phi = \varphi$. Equivalently,
\be
X^1 = R(z,\varphi)\, \cos\varphi\, ,\qquad X^2 = R(z,\varphi)\, \sin\varphi
\, ,\qquad
X^3 =z\,
\ee
We now have
\be
0=\bfn X^1 \times \bfn X^2 = RR_z\, .
\ee
In other words, the vanishing of ${\cal A}$ implies that $R_z=0$
(since $R=0$ would imply a collapsed D2-brane of zero area). This leads to
\bea
a_0 &=& \B\, , \nonumber\\
a_1 &=& E_z\left(R_\varphi\cos\varphi -R\sin\varphi\right) -
i \left(R_\varphi \sin\varphi + R\cos\varphi\right) \, , \nonumber\\
a_2 &=& E_z\left(R_\varphi\sin\varphi + R\cos\varphi\right) +
 i\left(R_\varphi\cos\varphi -R\sin\varphi\right)\, , \nonumber\\
 a_3 &=& -E_\varphi\, ,
\eea
and hence to
\be
\Delta^2 = \left(\B+E_\varphi\right)\left(\B - E_\varphi\right)
+ \left(1-E_z^2\right)\left(R^2+R_\varphi^2\right)\, .
\ee

Recalling that the conditions for preservation of 1/4 supersymmetry are
those in (\ref{onetwo}), we now find that they become
\be
\label{cw}
\B+ E_\varphi = c\Delta\, , \qquad
\left(E_z-1\right) e^{i\varphi}\left(R_\varphi + iR\right) = w \Delta\, ,
\ee
for real constant $c$ and complex constant $w$. In addition, we must
take into account the Bianchi identity
\be
\partial_\varphi E_z = \partial_z E_\varphi
\ee
and the Gauss law constraint
\be
\left(R^2+ R_\varphi^2\right) \partial_z\left(E_z/\Delta\right) +
\partial_\varphi\left(E_\varphi/\Delta\right) =0\, .
\ee

Let us first consider the cases for which $w$ in (\ref{cw}) is non-zero.
In this case the ratio of the real and imaginary parts of the second of
the supersymmetry conditions (\ref{cw}) yields a differential equation
for $R$ that has the solution
\be
{\cal I}m\left(\bar w e^{i\varphi}\right) R =R_0
\ee
for some constant $R_0$. This is a line in the $X^1$-$X^2$ plane parametrized
by $\varphi$, so the D2-brane geometry is planar. Further analysis shows that
the BI fields are constant in local cartesian coordinates, so $w\ne 0$
leads back to the 1/2 supersymmetric planar D2-branes already discussed.

The interesting case is therefore $w=0$, in which case supersymmetry
requires $E_z=1$, and hence
\be
\Delta^2 = \left(\B-E_\varphi\right)\left(\B+E_\varphi\right)
= c\left(\B-E_\varphi\right)\Delta\, .
\ee
This means that $\B$ and $E_\varphi$ are both proportional to $\Delta$
and hence to each other. As $\B$ cannot vanish (for positivity of $\Delta^2$),
we may write
\be
E_\varphi = \beta\, \B\, ,\qquad \beta = \frac{c^2 -1}{c^2 +1}\, ,
\ee
and hence
\be
\Delta = \sqrt{1-\beta^2}\, \B\, .
\ee
The remaining condition for supersymmetry ($\B+E_\varphi =c\Delta$) is now
an identity. Since $E_\varphi \propto \Delta$, the Gauss law states that
$\Delta$ is a function only of $\varphi$, which means that both $E_\varphi$
and $\B$ are also functions only of $\varphi$, since they are proportional
and their sum is proportional to $\Delta$.

We have now found 1/4 supersymmetric
D2-branes that are $z$-independent and hence invariant under translations along
the $Z$-axis. In fact, we have recovered the supertubes ($\beta=0$) and twisted
supertubes ($\beta\ne0$) of section \ref{sec:supert}. In that case we
found that $E_z=\pm 1$ but the $E_z=-1$ case is obtained from the $E_z=1$
case by a rotation, and in this section we effectively used the rotational
invariance to arrange for $E_z=1$ rather than $E_z=-1$.

\subsection{${\cal A}\ne0$}

When ${\cal A}\neq0$, we may choose
\be\label{fixgauge}
X^1=\sigma^1\, ,\qquad X^2=\sigma^2\, \qquad X^3= Z(\sigma^1,\sigma^2)
\ee
In this gauge, $\Delta = 1/|{\cal A}|$, and (\ref{onetwo}) is equivalent to
\be
\label{EB}
E= \bfn Z + {\bf k}\, ,\qquad \B = B_0+ {\bf k} \times \bfn Z\, ,
\ee
for constant $B_0$ and constant 2-vector ${\bf k}$. The worldspace metric is
\be
h_{ij} = \delta_{ij} + \partial_i Z \partial_j Z\, ,
\ee
from which one deduces, according to (\ref{Gauss}) and (\ref{displace}),
the Gauss law
\be
\label{genGauss}
\nabla^2 Z = \bfn Z \times \bfn \left({\bf k}\times \bfn Z\right)\, ,
\ee
so that $Z$ is harmonic when ${\bf k}={\bf 0}$. To make contact with
section \ref{sec:original}
we define new coordinates $(z,\varphi)$ by
\be
\sigma^1 = R\cos\varphi \, , \qquad \sigma^2 = R \sin\varphi\, ,
\ee
where the function $R(z,\varphi)$ is determined implicitly by the requirement that
\be
Z(\sigma^1, \sigma^2)= z\, .
\ee
The Laplace equation for $Z$ is now equivalent to the differential
equation (\ref{GaussR}) for $R$, and the BI field-strength 2-form is
\be
F= dz\wedge dt + B_0RR_z dz\wedge d\varphi\, .
\ee
We have therefore recovered the superfunnels, and other `supershapes' of
section \ref{sec:original}.

It remains to analyse the case of non-zero ${\bf k}$ or, equivalently,
non-zero $k\equiv |{\bf k}|$.
If the expressions (\ref{EB}) are used to compute $\Delta$ as given
in (\ref{delHG}) in terms of $\bfn Z$, then one finds that $\Delta$ is
a constant, as required, if and only if
\be\label{zeq}
B_0 \, {\bf k}\times \bfn Z - {\bf k}\cdot \bfn Z = \ell
\ee
for some constant $\ell$, which is not independent of those already introduced since
\be
\Delta^2 = B_0^2 -\left(1+k^2\right) + 2\ell\, .
\ee
The solution of this equation is
\be\label{ZW}
Z = W(z_+) + {\ell\over k \sqrt{1+B^2_0}}\, z_-\,,
\ee
where
\bea
 z_+ &=&{1\over k \sqrt{1+B^2_0}}
\left[ \left(B_0 k_1 -k_2\right) \sigma^1 + \left(B_0 k_2 + k_1\right)
\sigma^2\right]\, ,\nn
z_- &=&{1\over k\sqrt{1+B^2_0}}
\left[ -\left(B_0 k_2 +k_1\right) \sigma^1 + \left(B_0 k_1 - k_2\right)
\sigma^2\right]\,.
\eea
The Gauss law constraint (\ref{genGauss}) now yields
\be
\left(1+B_0^2 +\ell \right) W'{}' =0 \, .
\ee
The term in parentheses cannot vanish for real $\Delta$, so we conclude that
$W'{}'=0$, and hence that $W$ is a linear function of $z_+$. Thus,
\be
Z= Z_0 + Cz_+ + {\ell\over k \sqrt{1+B^2_0}}\, z_-\,,
\ee
where $Z_0$ and $C$ are constants. The geometry is therefore planar.
Moreover, the BI fields are constant, so that these configurations with
$k\ne0$ are actually just the 1/2 supersymmetric planar D2-branes that
we already know about.

\section{D2 in 4D}
\label{sec:4D}

There is a further class of 1/4 supersymmetric configuration of D2-branes
that is suggested by the well-known 1/4 supersymmetric K\"ahler calibrated
M2-branes. The latter is a configuration that has two asymptotic planes and
may therefore be interpreted as an intersection of two M2-branes,
although the intersection is non-singular so the configuration can be found
as a solution of the worldvolume equations of motion for a single
supermembrane. It is obvious that dimensional
reduction of such a configuration could lead (depending on its orientation)
to a similar D2-brane configuration. The geometry is intrinsically
4-dimensional, which explains why we did not see it in our previous
exhaustive analysis of D2-branes in 3D. As we now show, a novelty of
the D2-brane version of this `intersecting brane' configuration is the
possibility, consistent with 1/4 supersymmetry, of a superposed
uniform BI magnetic field.

As we expect the geometry to be 4-dimensional, we set $X^5=X^6=X^7=X^8=X^9=0$.
We fix the worldspace parametrization invariance by setting
\be
X^3= \sigma^1, \qquad X^4= \sigma^2\, ,
\ee
leaving $(X^1,X^2)$ as the surviving worldvolume scalar fields. We now find that
\be
\det h=1+ \left|\bfn X^1\right|^2 + \left|\bfn X^2\right|^2
+ \left(\bfn X^1 \times \bfn X^2\right)^2\, .
\ee
We will set the BI electric field to zero, so the Gauss law is satisfied
trivially, and
\be
\Delta^2 = \det h +\B^2\, .
\ee
The kappa-symmetry matrix of (\ref{redgam}) is now
\bea
\Gamma &=& \Delta^{-1} \bigg[\Gamma_{T12} + \Gamma_{T13}\,
\partial_2 X^1 - \Gamma_{T24}\, \partial_1X^2 -\Gamma_{T23}\, \partial_1X^1
+ \Gamma_{T14}\, \partial_2 X^2 \nonumber\\
&&\qquad\qquad +\ \Gamma_{T34}\left(\bfn X^1\times \bfn X^2\right) +
\Gamma_{T\natural}\, \B \bigg]\, .
\eea
Imposing the conditions
\be\label{1234}
\Gamma_{1234}\epsilon=\epsilon\, ,
\ee
and the Cauchy-Riemann (CR) equations
\be\label{CR}
\partial_1X^1= -\partial_2X^2\, ,\qquad \partial_1X^2 =\partial_2 X^1\, ,
\ee
one finds that
\be
\sqrt{\det h} = 1+ \frac{1}{2}\left|\bfn X^1\right|^2
+ \frac{1}{2} \left|\bfn X^2\right|^2\, ,
\ee
and the supersymmetry condition $\Gamma\epsilon=\epsilon$ reduces to
\be\label{dyonint}
\left[\sqrt{\det h}\, \Gamma_{T12} +\B\, \Gamma_{T\natural}\right] \epsilon
=\Delta\, \epsilon\, .
\ee

Provided that the magnetic field $\B$ is uniform, which means that
\be\label{Bden}
\B = B\sqrt{\det h}\, ,
\ee
for some constant $B$, the two constraints (\ref{1234}) and (\ref{dyonint})
are compatible and imply preservation of 1/4 supersymmetry. The CR equations
imply that the complex field ${\cal Z}=X^1+iX^2$ is a holomorphic function of
the complex worldspace coordinate $\zeta= \sigma^1-i\sigma^2$. Equivalently,
one has $f({\cal Z},\zeta)=0$ for holomorphic function $f$ of the two complex
variables $({\cal Z},\zeta)$; this implies that the D2-brane is K\"ahler
calibrated. The novelty here is the additional uniform magnetic field density.

\section{General cross-sections}
\label{sec:general}

We know from \cite{Mateos:2001pi} that the most general cross section of
a supertube is a an arbitrary curve in the 8-dimensional space transverse
to the axis of the tube. An obvious question is whether this
result generalizes to twisted supertubes and superfunnels. We now aim to
answer this question, using the cartesian coordinate approach
of \cite{Mateos:2001pi}. We first set
\be
X^I= (Z, Y^A)\, \qquad (A=1,\dots,8)\, ,
\ee
where the $Z$-axis will be the axis of the supertube or superfunnel.
Then we observe that
\bea
\Delta^2 &=& \left[(\bfn Z +{\bf E}) \times \bfn {\vec Y}\right] \cdot
\left[(\bfn Z -{\bf E})\times \bfn {\vec Y}\right] - \left({\bf E}\times
\bfn Z\right)^2 \nonumber\\
&&\ + \sum _{A>B} \left(\bfn Y^A\times \bfn Y^B\right)\left(\bfn Y^A\times
\bfn Y^B\right) +\B^2\, ,
\eea
where $\bfn Y^A$ are the components of the 8-vector $\bfn{\vec Y}$ and
the dot product is the standard Euclidean inner product on $\bE^8$.
In the same notation we have
\bea\label{delgamma}
\Gamma &=& \Delta^{-1}\bigg[\left(\bfn Z\times \bfn Y^A\right)\Gamma_{TZA} +
{1\over2}\left(\bfn Y^A \times \bfn Y^B\right) \Gamma_{TAB} \nonumber\\
&& \ +\, \left({\bf E}\times \bfn Z\right)\Gamma_{Z\natural}
+ \left({\bf E}\times \bfn Y^A\right)\Gamma_{A\natural}
+ \B\Gamma_{T\natural}\, \bigg].
\eea

For either superfunnels or supertubes, twisted or otherwise,
we impose the constraint
\be
\label{firstcon}
\Gamma_{TZ\natural}\epsilon=\mp\epsilon\, ,
\ee
which allows us to rewrite (\ref{delgamma}) as
\bea\label{delgamma2}
\Gamma &=&\Delta^{-1}\bigg[ \left({\bf E} \mp \bfn Z\right) \times \bfn Y^A\,
\Gamma_{A\natural}
+ \left({\bf E}\times \bfn Z\right)\Gamma_{Z\natural} \nonumber\\
&& \ +\, \frac{1}{2}\left(\bfn Y^A \times \bfn Y^B\right) \Gamma_{TAB}
+ \B\Gamma_{T\natural}\, \bigg].
\eea
To proceed, we the solve the Bianchi identity (\ref{curlE}) by setting
\be
{\bf E} = \bfn V
\ee
for some electric potential function $V(x,\varphi)$. As this function depends
on $E_\varphi$, we consider first the cases in which $E_\varphi=0$,
which are the superfunnels and untwisted supertubes.

\subsection{Superfunnels}

Let us choose
\be
V= \pm Z
\ee
In this case the supersymmetry preservation condition becomes
\be\label{susypres2}
\left[{1\over2}\left(\bfn Y^A \times \bfn Y^B\right) \Gamma_T\Gamma_{AB}
+ \B\Gamma_{T\natural} -\Delta\right] \epsilon =0\, ,
\ee
where now
\be
\Delta^2 = \sum _{A>B} \left(\bfn Y^A\times \bfn Y^B\right)\left(\bfn Y^A
\times \bfn Y^B\right) +\B^2\, .
\ee

If there are to be no further constraints on $\epsilon$, we must have
\be\label{Omega}
\bfn Y^A \times \bfn Y^B = \Delta\, \Omega^{AB}
\ee
for some {\it constant} antisymmetric $8\times 8$ matrix $\Omega$.
One possibility is $\Omega=0$, which arises when $\partial_x{\vec Y}=0$. This
leads to the standard (untwisted) supertube with arbitrary cross-section.
We skip the details since the untwisted supertube may be considered as
a special (zero-twist) case of the twisted supertube, to be discussed
in the following subsection. However, for purposes of comparison
with the superfunnel, we observe that when $\partial_x{\vec Y}=0$
the equation (\ref{Omega}) places no restriction on the $\partial_\varphi{\vec Y}$,
and this is what allows the supertube cross-section to be an arbitrary
curve in the transverse 8-space. As we shall now see, the situation is
quite different when $\Omega$ is non-zero.

Given that $\Omega$ is non-zero, (\ref{Omega}) implies
(i) that $\Delta$ is a constant in the gauge
\be\label{gaugefixY}
Y^1=\sigma^1\, ,\qquad Y^2=\sigma^2\, ,
\ee
and (ii) that the projection of the D2-brane geometry on the 8-dimensional
Euclidean space with coordinates $Y$ is a plane, which we may orient
such that the only non-zero entries of $\Omega$ are
$\Omega^{12}=-\Omega^{21}= 1/\Delta$. We then have $\Delta^2= 1+\B^2$ but
since $\Delta$ is a constant, we have $\B=B$ for some constant $B$.
The supersymmetry preservation condition (\ref{susypres2}) now becomes
\be
\Gamma_T\left[\Gamma_{12} + B\Gamma_\natural\right] \epsilon
= \Delta \epsilon\, ,\qquad
\Delta= \sqrt{1+ B^2}\, .
\ee
By itself, this condition would imply preservation of 1/2 supersymmetry,
as would (\ref{firstcon}). These two constraints on $\epsilon$ are
compatible because $\Gamma_{TZ\natural}$ commutes with both
$\Gamma_T\Gamma_{12}$ and $\Gamma_{T\natural}$, and the two together
imply preservation of 1/4 supersymmetry.

To summarize, we have now shown that any configuration with constant $\B$,
and electric field ${\bf E}= \pm \bfn Z$, preserves 1/4 supersymmetry.
However, we have still to impose the Gauss law constraint.
Since ${\cal D}_i = \pm \Delta^{-1} \partial_i Z$ and $\Delta$ is constant,
the Gauss law is
\be
\nabla^2 Z=0\, .
\ee
We have now recovered our earlier result for superfunnels, but now
we have seen that the superfunnel cross section is necessarily planar.

\subsection{Twisted supertubes}

For convenience of comparison with our analysis of section \ref{sec:original},
we make the notational change $(\sigma^1,\sigma^2) = (z, \varphi)$.
Returning to (\ref{delgamma2}) we now partially fix the worldspace
diffeomorphisms by the gauge choice
\be
Z= z \, ,
\ee
and we further assume that all other worldvolume fields are $z$-independent.
We then have a D2-brane configuration that is invariant under translations
along the $Z$-axis. Also, (\ref{delgamma2}) simplifies to
\be\label{delgamma21}
\Gamma= \Delta^{-1}\bigg[\partial_\varphi Y^A \Gamma_{A TZ}
+ E_z\partial_\varphi Y^A \Gamma_{A\natural}- E_\varphi\Gamma_{Z\natural}
+ \B\Gamma_{T\natural}\bigg]\, ,
\ee
where
\be
\Delta^2 = \left(1-E_z^2\right) |\partial_\varphi{\vec Y}|^2 -E_\varphi^2
+ \B^2\, .
\ee
The twisted supertubes are now found by setting
\be
E_z=\pm1\, , \qquad \Gamma_{TZ\natural}\epsilon = \mp \epsilon\, .
\ee
The Gauss law is now an identity, as is the Bianchi identity, and
the supersymmetry preservation condition is
\be
\left(\B \Gamma_{T\natural} - E_\varphi \Gamma_{Z\natural} \right)\epsilon
= \Delta \epsilon,
\qquad \Delta^2 = \B^2-E_\varphi^2\, .
\ee
For this to be satisfied without further constraint on $\epsilon$, we require that
\be
E_\varphi = \beta \B \, , \qquad \beta^2<1\, ,
\ee
for some constant $\beta$, as we found in (\ref{EBvarphi}).
Given the above electric field components, we see
that the electric potential is
\be
V(z,\varphi)= \pm z + \beta \int^\varphi d\varphi' \B(\varphi')\, .
\ee
As expected, this coincides with the choice $V=\pm Z$ of the previous
subsection when $\beta=0$ (since we are now working in the gauge $Z=z$).
The supersymmetry preservation condition is now
\be
\frac{1}{\sqrt{1-\beta^2}} \left(\Gamma_T - \beta\Gamma_Z\right)
\Gamma_\natural \, \epsilon =\epsilon\, .
\ee
As both $\Gamma_{T\natural}$ and $\Gamma_{Z\natural}$ commute with
$\Gamma_{TZ\natural}$, this constraint is compatible with (\ref{firstcon})
and the two together imply preservation of 1/4 supersymmetry.

The above result can be summarized as follows. The static D2-brane
configuration with
\be
Z=z\, ,\qquad {\vec Y}= {\vec Y}(\varphi)\, ,
\ee
and
\be
F= \pm dt\wedge dz + q\B(\varphi)\, dt\wedge d\varphi + \B(\varphi)\,
dz\wedge d\varphi\, ,
\ee
preserves 1/4 supersymmetry for arbitrary functions ${\vec Y}(\varphi)$,
and arbitrary positive function $\B(\varphi)$. Although it is not necessary,
we may suppose that $\varphi$ is periodically identified, such that
$\varphi\sim \varphi+2\pi$ without loss of generality, and
in this case we have a tubular configuration, translationally invariant
along the $Z$-axis, and with a cross-section determined by an arbitrary
closed curve in the transverse 8-dimensional space. For $\beta=0$ this
is the general D2-brane supertube of \cite{Mateos:2001pi}, for which
the electric field lines are parallel to the $Z$-axis. Note that
although the function $\B(\varphi)$ is arbitrary, we have still to fix
the $\varphi$-reparametrization invariance. As $\B>0$, this can be done
by setting $\B=B$ for some positive constant $B$.
Since $B= \oint\! d\varphi\B(\varphi)$, which is reparametrization
invariant, different choices of $B$ represent distinct configurations.

When $\beta\ne0$, we have a new class of 1/4 supersymmetric D2-branes for
which the electric field lines are at a non-zero angle to the $Z$-axis
and therefore twist around it. These could be called ``twisted supertubes''
but they are actually just standard supertubes boosted in the $Z$ direction.
Note that $\Delta= \B\sqrt{1-\beta^2}$ is boost invariant, so a boost from
rest increases $\B$ and hence the energy density. In fact, the energy density is
\be
{\cal H} = \frac{|\partial_\varphi {\vec Y}|^2 +B^2}{B\sqrt{1-\beta^2}} =
\frac{1}{\sqrt{1-\beta^2}}\left[\left|{\cal D}\right|_{\beta=0} + B\right] \, ,
\ee
as expected for a boost of the supertube with velocity $\beta$.

\section{Discussion}

The work reported here began with the aim of clarifying some aspects of
the supersymmetry preservation condition for solutions of the
Dirac-Born-Infeld equations for a D2-brane in a 3-dimensional Euclidean space.
We have presented an exhaustive analysis of the conditions for supersymmetry
preservation under this restriction and one result of this analysis is
that all supersymmetric D2-branes of this type preserve either 1/2 or
1/4 supersymmetry.

The general 1/2 supersymmetric solution includes the obvious planar D2-brane
vacuum but also includes other planar solutions with constant Born-Infeld fields.
All these planar D2-branes lift to a planar M2-brane in M-theory, but the
identification of the M-theory circle coordinate implies a finer classification
of 1/2 supersymmetric D2-branes than one has for planar M2-branes in the
11-dimensional Minkowski vacuum. This classification arises by considering
the velocity of the intersection of the M2-brane with what we have called an
`ether'-brane, and a `null intersection' leads to a `dyonic' D2-brane.

In the case of 1/4 supersymmetry, we found a new class of `twisted' supertubes,
with electric field lines that twist around the tube. These can be understood
as supertubes boosted along the direction of translational invariance.
The boost invariance in this direction is broken by the magnetic field density,
which is why a boost generates a new solution. We also found `superfunnels',
which are tubular configurations with arbitrary planar cross section, with
a scale that varies (exponentially, on average) along the tube. Various other
1/4 supersymmetric `supershapes' were found to be possible, including
asymptotically planar D2-branes.

We have also analysed the conditions for 1/2 and 1/4 supersymmetry without
the restriction to an embedding in 3-dimensional space. In four space dimensions
there are K\"ahler-calibrated minimal surfaces that are 1/4 supersymmetric solutions
of the DBI equations with vanishing BI fields, and we found a generalization of
this that allows for a uniform magnetic field. Our analysis was not exhaustive
so there may be other types of 1/4 supersymmetric time-independent solutions
for which the embedding space has more than three dimensions, but we suspect
that nothing essentially new is possible.

What is known is that supertubes may have a cross-section that is an arbitrary
curve in the 8-dimensional space transverse to the tube, and we have shown that
the same applies to twisted supertubes. In contrast, the cross-section of
a superfunnel was found to be necessarily planar. This difference might seem
surprising but the translational invariance of supertubes makes possible
string-theory dual configurations in which the arbitrary cross-section
becomes an arbitrary wave profile, and this `explanation' is not available
to superfunnels.

It would be interesting to extend the considerations of this paper to lower
fractions of supersymmetry, and also to extend the analysis to D$p$-branes
with $p>2$. For example, for $D5$-branes one expects the BPS conditions
classified in \cite{Bak:2002aq} for gauge fields in a flat six-dimensional
spacetime to be relevant. We leave this to future investigations.

\bigskip
\noindent
{\bf Acknowledgements.}
DB would like to thank the Particle Theory Group of
University of Washington for the warm hospitality.
This work was initiated while NO was visiting DAMTP at Cambridge University,
and he also acknowledges its warm hospitality.
The work of DB is supported in part by
KOSEF ABRL R14-2003-012-01002-0 and KOSEF SRC CQUeST R11-2005-021.
The work of NO was supported in part by the Grant-in-Aid for
Scientific Research Fund of the JSPS Nos. 16540250 and 06042.
PKT thanks the EPSRC for financial support.

\end{document}